# User Influence Analysis Based on Blogs*

Xiang LIU, Yan JIA, Rong JIANG and Yong QUAN
*Dept. of Computer, National University of Defense Technology,*
*Changsha, 410073, China*
*E-mail: xiang_liu2013@163.com*
*www.nudt.edu.cn*

Abstract: Rumor and word of mouth spread in the same speed for the highway of information diffusion in the age of the internet. Social network plays quite an important role in the huge internet. Nowadays Social networks have become indispensable in our lives, especially for the government and the enterprise. Social network becomes a complex information diffusion network with user working as the node and the relationships between users working as the vehicle. In this paper we come up with three kinds of algorithm computing user influence based on the behavior of user's forwarding micro blog and the symbol of @ in the microblog. We evaluate the effectiveness of the algorithm by the method of comparing the results of our work and the train data in the dataset and in the end it prove that our algorithm works good.

## 1. Introduction

With the continuous development of network technology, network shows more and more superiority in people's lives. Social networks are increasingly becoming an important part of people's daily lives. Through micro blog, we can get to know what our good friends was doing, are doing and what they will do. We can see some of the current hot information through the net, such as a star's willing to come to your homeland to host a concert, a movie's being released before long and so on. Micro blog has indeed become indispensable in our lives. In addition, micro blog also plays another important role: product promotion, viral marketing. We know that with the continuous development of network technology, the advertising industry has been transferred from newspapers, radio, television and other traditional media to the internet. Even some of the big events are closely related to the micro blog network. Such as the U.S.

* This work is supported by national key fundamental research and development program (973 program) No.2013CB329600,2013CB329604 and National Natural Science Foundation of China No.61372191.

presidential election in 2008, MySpace and Facebook, especially Facebook play a very important role. The president has established a good personal image through communication and interaction with supporters on the Internet. Of course there are some negative examples, for example, some criminals use social networks to spread false information, spread rumors. This requires our government to strengthen public opinion monitoring to better ensure the healthy development of the internet. Strengthen the Internet user behavior monitoring, we can accurately grasp the development trend of social networks.

There is too much user information in the Social network that this paper only take forwarding micro blog and @ relationship information in the social network for user influence analysis. When we browsing other people's blogs, we found that some of the micro blog is useful to us or is interesting to us, thus we will forward his blog; likewise when our blog is useful to other users, they will forward our blogs too. For some well-known big V, there will be a lot of micro blog users who forward their micro blog; relatively speaking, we common users will have much less people forward our blogs. In these cases, we believe that the number of users who have more blogs forwarded are more influential than the users who have less users forwarding their blogs. For @ information, to a certain extent, it is more similar with the forwarding behavior; we will carry out a detailed account in this text. In this paper, the third kind of social network user influence calculation method is based on the forwarding level. That is to say, for the case that when the user B forwards a blog of user A, user C forwards the blog of user B, it forms a forwarding chain structure. with social network structure more and more complex, it will form a complex forwarding structure named forwarding tree. In this paper, we will figure out user influence by the forwarding tree. Overall this paper works on the relationship between the forwarding behavior and the content of blogs information to calculate the user influence. Our result show that our algorithm performs rather good.

The rest of this paper organized as follows: the second chapter introduces the previous related work; the third chapter introduces the data sets we use; the fourth chapter introduces our three algorithms: forwarding, @ and forwarding hierarchy; the fifth chapter presents our experimental results; the sixth chapter is the evaluation of the effectiveness of the algorithm; the seventh chapter is the conclusion.

## 2. Related Work

In recent years, with the study of user behavior in social network deeper and deeper, various methods of research about user influence studying from different all kinds of perspectives emerge in an endless stream. We carried out the analysis of user influence from different angles, and launched it in the view of their own points. This plays a great role in guiding us to understand the user behavior in social network.

Daniel[1] et al. proposes IP algorithm, based on research about influence and passivity. They have got analytical factors by analyzing the behavior of users to click on the URL of others, which will be helpful to future analysis. Adrien[4] et al. through the complex calculation of the way of information transmition, have had put forward a practical model which can be used to guide the present network media to spread information much better and faster. This study can help us understand the present work better, and propose the improvement measures to the present work. What is more, there are some topic-themed anslysis of the user influence. For example, Tang[5] et al. has put forward a topical affinity propagation to simulate the topic level user behavior to analyze the user’s influence. TAP model works on the MapReduce model proposed a series of efficient distributed algorithm and show its advantages by the analysis of the effectiveness. Zhang[6] et al. thinks the user behavior is mainly affected by the most intimate friends in his ego network. Based on this understanding, their social influence positioning technology can be a good predictor of user forwarding behavior. From a dynamic point of view, Valerio[3] et al. has analyzed the changes of the user’s behavior in social network , and think it better to use OSN model to analyze the change rules of user’s behavior. It also can make them to understand the network actions better in the past by OSN model, and also it is better to predict the dynamic changes of the user’s behavior in the future.

## 3. Dataset

We have used the crawler technology to crawl information of users from Sina micro blog continuously, including the data from 2014 March to May. This dataset is used as a prediction dataset. First of all, we need to clean up the data, and remove the accounts which doesn't conform to our analysis. Considering that crawling too much data will lead to cost too much time, and also make analysis quite too slow, so we have selected 200 thousand user's information, each user has blog of on average of 200 blogs, a total of 40 million of articles in the blog. Dataset's format is shown as in table 1:

Table 1. Data Format

| Field No. | 0 | 1 | 2 | 3 | 4 | 5 | 6 |
|---|---|---|---|---|---|---|---|
| Field Name | Did | Uid | Username | Timestamp | Uid | Did | content |

Among them: Uid refers to the user ID ;Did refers to the document ID, and it is the id of the user's blog; Username refers to the user's nickname; Timestamp refers to the time when post his blog; Content refers to the blog content. When we do the data analysis, we will need some information of the blog, such as the information of @ . Therefore, the Content section is an important part of our data set.

Besides, in order to carry out data analysis, we also have chosen some users of Sina micro blogging celebrity to collect data, and probably get 50 users' information. This data format is same as the user data format is same as table 1. This data set regarded as training dataset carries on to forecast the predicting dataset.

## 4. Methods

Analyzing the influence among the users, blog plays an important role. There are a lot of papers on the study of blog. This article measures the influence between users only from the blog forwarding and the @ information.

### 4.1. *Forward*

Every blogger may have many posts, each blog is likely to be forwarded by the blogger's friends and such formats of a forwarding behavior.

Based on past experience, we are likely to boil the user's forwarding other people's blog down to two main reasons:

i)This blog has attracted the user who forwarding the blog by the content of the blog; that is to say, the blogger have a certain influence on the user who forwarding the blog.

ii)That is online water army. There are quite a lot of research on exposure of online water army now, and it already has so much mature technology that We don't make a detailed explanation here; and the data set has been cleared about the online water army.

From the above information, I believe that you will have a general understanding of this article about the forwarding blogs; and the following will be described in detail about the forwarding algorithm.

We define V as the user space, including all users in the dataset of this text, define the user i as Vi. Define D as all the blog space, and the $j_{th}$ blog of user I names $D_{ij}$. The blog posts of user i as $D_{vi}$.

We have analyzed that the user's forwarding other people's blog just because the user is affected by others, which can reflect the user's influence of the user on other users who has a direct relationship. However, if we just take the number of forwarding behavior to calculate the influence is obviously not enough, we also need to consider the total number of blogs of the user. This is because, although there may be a certain number of users forwarding his blogs, if the number of blogs of the user is quite too small, it will seriously harm the blogger's influence and his fans will decline. This also shows that the blogger does not pay much attention to its fans. Therefore, when we calculate the user's influence on the forwarding relationship, it also need to be reflected in the total number of the users' blogs.

By the above statement, we get the formula for calculating the user's influence by the amount of his blogs which are forwarded and the total number of blogs:

$$I(v_i) = \frac{F(vi)}{D(vi)}, \tag{1}$$

Through this formula, we can figure out the user influence if we have the number of the number of a user's blogs forwarded and the total number of the user's blogs.

### 4.2. *Symbol of @*

For users who are often or less frequently (but occasionally) used micro blog, there is a sign that everyone is very familiar with which is @. When we are posting a blog, adding a @ symbol in the micro blog, we can make the users who was @ to see our blogs. Of course, this is also divided into several cases:

i) who was @ is the famous big V. And they are people who we seldom communicate with in our daily lives. We @ them just because we have question to consult them or we want to make them see some of our views.

ii) who ware @ are our good friends. They may have minimal influence in the social network, but who we are good friends! We @ each other just to share a thing, or a news, or funny story. Only this and no more just!

Of course, the @ relationship is more complex than the forwarding relationship.

For each micro blog, it is likely consist @ relationship, and for the bloggers, he or she will have many @ relationship, analysis of the number of these @ relationship in coarse granularity, we can draw the following conclusions:

i) The user who @ many people and also many people will @ him. These people are usually more active, and we think that they are welcome.

ii) The user who seldom @ others but there are many people who will @ him. Such people in the social network is not too much. They are often some organizations or medias or individuals which Have high reputation and popularity, such as China Daily, CCTV, Ma Yun, etc..

iii) The user often @ other people, but there is few people will @ him. This situation is not very common. In the social network, some users are very active, but the blog of such users may be too shallow, it is difficult to resonate around other people, so these posts are lack of other people's attention.

iv) Still there are people who seldom @ other people and there are few people will @ him. This person may be online water army, may be some interference users or probably these people are really less active in the micro blog. We may have heard about some of these people that who seldom post blogs, but often read other people's blogs.

See from above, in fact, there may be some people still feel forward and @ relationships are the same, in fact, these two points are different. The difference is in that each blogs, users can @ a number of users, which led to the number of users bigger than the number of users forwarding blogs. We know that when the user A @ other people B, B will has a greater chance to read the blog which consist of the symbol @. As long as he read the blog there will be a certain degree of probability of being influenced. So, in a sense, the @ relationship has more power than the forwarding relationship.

Therefore, through the above elaboration, to calculate the influence of the user, intuitive, we may draw the following formula:

$$I(v_i) = a \times I(i) + b \times I(ii) + c \times I(iii) + d \times I(iv). \qquad (2)$$

Among them, I(i), I(ii), I(iii), I(iv) are the influence of the above four kinds of users, a, b, c, d corresponding to the influence coefficients. But after careful analysis, we found a problem: for the user vi, those users who @ vi is equivalent of vi’s fans. @ behavior can be understood as that the user has already be influenced by vi (they @vi just because that they are concerned about vi, and this can be understood as having been affected by vi); and to those users who vi @ (of course, some of these users have @ VI already, here we should remove this part of the users), means that vi will be applied to the impact or trying to influence the users (however, whether he really be affected, there is a lot of uncertainty). This is the idea of the @ relationship to calculate the user's influence.

From the above analysis, we come to the conclusion that the influence should be included into two parts: the users that will be affected and the users who have already been affected:

$$I(v_j) = Ied(v_j) + c \times Iing(v_j). \tag{3}$$

Here, Ied represents the number of people influenced which is already confirmed.

Iing represents the number of people is to be influence. The parameter c represent influence coefficient. This parameter is used to deduce Iing from Ied.

Through the analysis of the data, we know that the user data which has been affected in the data set is easy to get. The problem we have to discuss is the acquisition of user data that is about to be affected.

After data analysis, we separate the dataset to four kinds of data:

i)The dataset of the users who @ user vi, named $R(vi)$ . We regard these users as the user vi has already influence.

ii)The dataset user vi @, this dataset names $S(vi)$ .

iii)The dataset is the users who already has @ interaction relationship with user vi, this dataset names $T(vi)$, it's the intersection of dataset $R(vi)$ and $S(vi)$ .

iv) Remove the dataset of $T(vi)$ from the dataset of the users vi @ ,and we name the left dataset $Q(vi)$ . That's to say $Q(vi) = S(vi) - T(vi)$ .

For further analysis, we found that i and ii can deduce iii, in other words i and ii are the past iii.

For iii, the possibility the user vi affect other users around him in the social networks is:

$$P(iii) = \frac{T(vi)}{S(vi)} = c \tag{4}$$

Then, we rank user influence of the @ behavior by the number the user influence, the formula of the number a user influence is:

$$I(vi) = R(vi) + c \times Q(vi) \tag{5}$$

### 4.3. *Forwarding hierarchy*

Searching in micro blog, we are very easy to find another problem, that for every blog post, it may be forwarded more than once, it may be forwarded by more than one person. What's more important is that when the user B's blog was forwarded by a user A, it is likely that there will be another user C forwarding B blog, there is still probability that a user D forwarding user C blog, and this forms a forwarding chain.

So for this kind of situation, if we just simply regard the forwarding behavior of user B forwarding user A's blog as the influence factor, obviously it weakened its influence. Therefore, here we have to talk about the problem lies in the forwarding chain composed by the forwarding relationship. And how shall we calculate the user's influence?

We know that not long ago Google has published an article of web page authority algorithm on the field of information retrieval named PageRank. They set score to the web pages about the importance based on PageRank algorithm; and finally based on this score they establish a certain rules to better meet user's search request of information. The forwarding chain of blogs here is similar to the web pages described in PageRank algorithm. So to get the score of the blogger we should add up all the score of each posts in the forwarding chain. There is a need to note that there are a number of blog posts, and there may be a number of blog posts in these posts which are forwarded more than once. Taking this into account, we should differentiate every score of the post and then add them to get the user influence of the specific blogger.

But we found that there are essential differences between the users of social networks and the web pages in PageRank. For example, in PageRank algorithm, for each page, the several links has the same weight. But for each common bloggers, every degree is a post. But the blog is very different, because bloggers even some of the big V, many of the blogs also are not forwarded by others. Although there are so many difference, the thought of the score of each page given by the previous page is still so us much favor. For each user B forwarding blogs of user A and then user C also forwarded the blog from the user B, but no matter whether user C forward blog from user A, the source of the blog is from A. So we believe that A's score is incline by one owing to C's forwarding blog. While for user B, C did not contribute to incline the score of B. (We insist that the size of a blogger's influence is determined by the number of times it is forwarded, but not as the fact that high quality website contributes high score in PageRank)

The dataset can be divided into two parts, one part is the list of blogs without being forwarded. For this part, the contribution of the blogs to the blogger is 0. That's to say these posts can't incline the score of the user. For the other part, it's consist of a list of blogs which is forwarded by others. For this part, every blog make the score of the blogger incline by one.

We can separate several forwarding trees from the source dataset. And these forwarding trees consist of the forwarding forest.

We name the forwarding tree(FT) as Ft(vi) and the forwarding tree is shown in Fig.1; and the forwarding forest(FF) include Ft(0),Ft(1),Ft(2),Ft(3)… and the forwarding forest is shown in Fig.2.

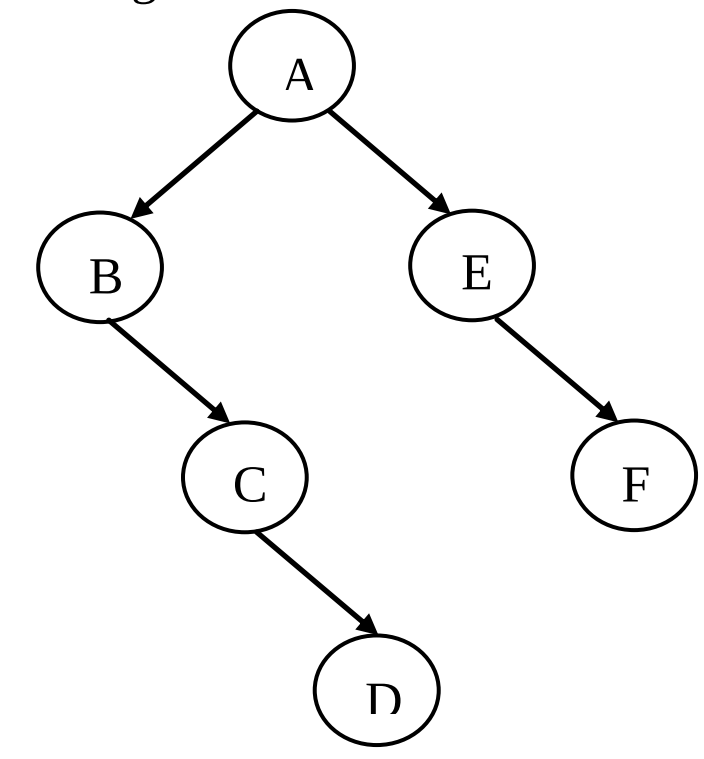

Fig.1. Forwarding Tree (FT)

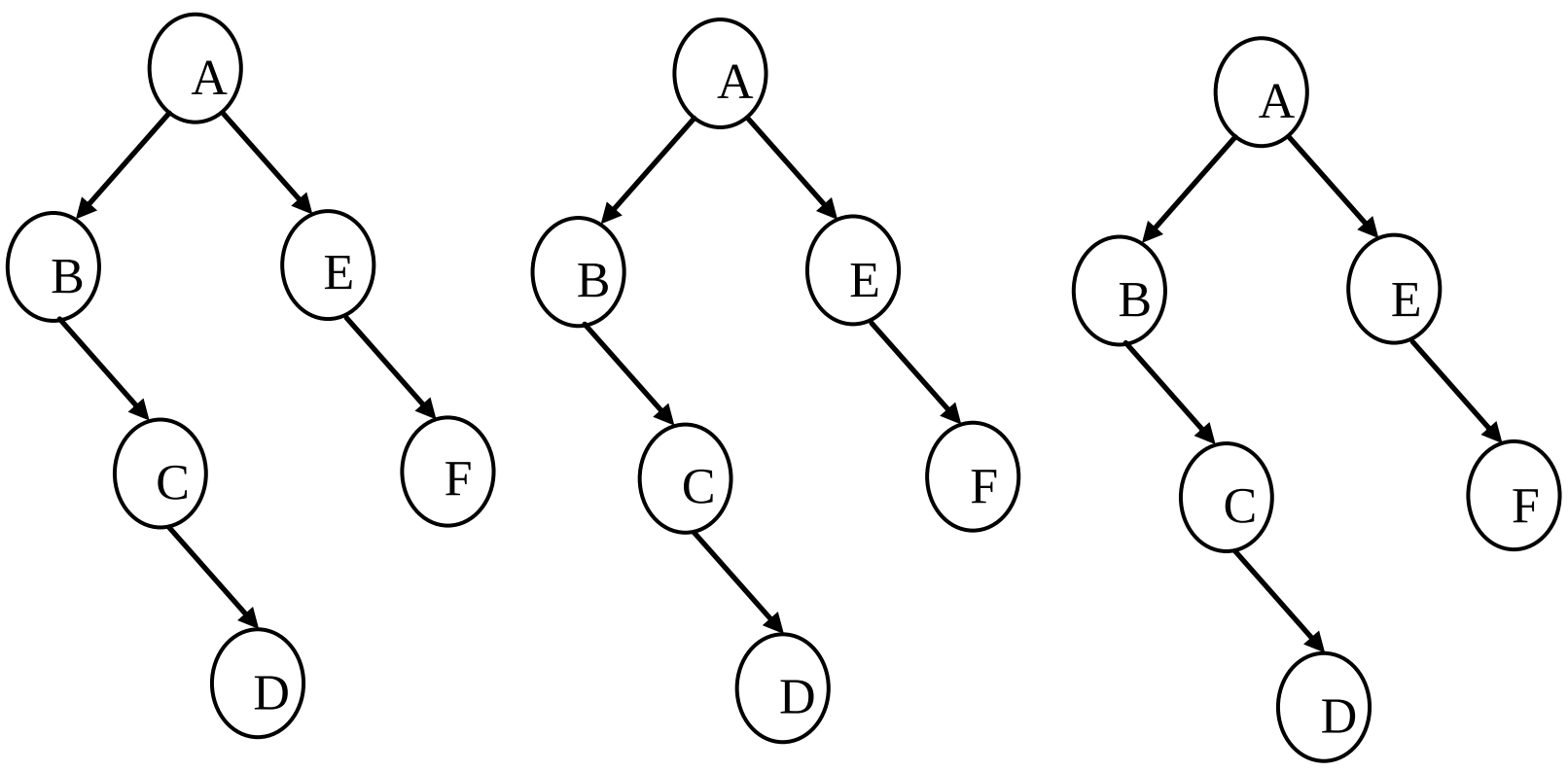

Fig.2. Forwarding Forest (FF)

Traversal tree structure of all the sub nodes we can get

$$\mathrm{Fscore}(i) = n \quad \text{(n is the number of all the sub nodes)}. \tag{6}$$

For the forwarding forest:

$$\mathrm{FScore}(vi) = \sum_{vi \in V} \mathrm{Fscore}(vi). \tag{7}$$

FScore(vi) is the sum of the contribution value of each node to the root node, and that is the score of each node vi. Obviously the sum of contribution of all other non forwarding relationships is

$$\sum_{\mathbf{Fscore}\,(vi)} 0 = 0 \ . \tag{8}$$

At this point, for all the nodes sorted by value, it is ranking of user influence.

## 5. Results

### 5.1. *Results of the three algorithms*

According to the analysis of the chapter 4, we design the corresponding algorithm, and finally get the results which is shown is table 2:

Table 2. Rank of User Influence by Forwarding Behavior and @ Behavior

| Forward | @ |
|---|---|
| 1708942053 | 1708942053 |
| 2090591961 | 2482557597 |
| 2547916923 | 2090591961 |
| 2482557597 | 1866402485 |
| 1866402485 | 1649155730 |
| 1649155730 | 2547916923 |
| 3249656920 | 1742703365 |
| 1652707015 | 1725876907 |

Notes: the number is Uid.

We can see the results of the two algorithms as shown above, and for the result of Forwarding Hierarchy is shown in table 3:

Table 3. Rank of User Influence by Forwarding Hierarchy

| Forwarding Hierarchy |
|---|
| 1708942053 |
| 2482557597 |
| 1866402485 |
| 1649155730 |
| 2547916923 |
| 2090591961 |
| 1725876907 |
| 1652707015 |

Notes: the number is Uid.

### 5.2. *Evalution*

For the data obtained in the last section, we compare the results obtained in the last section with the train data and find that forwarding hierarchy performs better and the other two perform alike. The three results consist of three vectors, which need to be standardized and then can be used for the calculation. The standard formula is as follows:

$$I(vi) = \frac{Rx(vi)}{\sum_{vj \in V} Rx(vj)}. \tag{9}$$

Ry and Rz is similar to Rx.

Then we have to carry out the similarity comparison of the three results, the method used here is cosine similarity, the formula is as follows:

$$Sim(A, B) = \cos\theta = \frac{A \cdot B}{|A| \cdot |B|}. \tag{10}$$

Through the above formula, first the standardization, and then calculate the similarity.

The cosine similarity calculation results is shown in table 4:

Table 4. Cosine Similarity of Three Algorithms

| Algorithm | Cosine Similarity |
|---|---|
| Forward & @ | 0.719 |
| @ & Forwarding Hierarchy | 0.522 |
| Forwarding Hierarchy & Forward | 0.499 |

By calculating the cosine similarity of the three algorithms, we can see that the overall results of the three algorithms are not bad, of which the result of forward and @ relationship based algorithms is close, but it is not too fit with the actual situation. The algorithm based on the forwarding layer fit more with the actual data.

We can see that, it’s not good enough to get the user influence just according to the forwarding behavior and the @ relationship. Because it just concerned about the user’s information and the user’s close friends; but we are more likely to get the user's actual influence from a more comprehensive network structure that is based on the forwarding hierarchy. Therefore, we can strengthen the detection of these more influential people ranked by forwarding level based algorithm. Only do like this can we better grasp the network development trends. And also we can be better to promote our products through these users.

## 6. Conclusions

Through this article, we can understand the great significance of the user influence in the social network to our society in many aspects. So the research on the user's influence is still in constant. Many experts and scholars have contributed a lot of excellent algorithms in this area, and make a great contribution to our government and enterprises. In this paper, we attempt to come up with a new algorithm on user influence based on the forwarding and the @ symbol in the microblog. We know that the PageRank algorithm is a web ranking algorithm, although there indeed is certain similarities between PageRank and social network users influence calculation, the social network has more of its own characteristics, we also proposed an influence algorithm based on the forwarding hierarchy.

In the next step, we will further calculate the user influence on a fine grain. Previous methods focus more attention on computing users' personal influence, the next step we will pay more attention to explore the influence between the users and bring corresponding algorithm. In order to grasp of the development of social networks more sensitively. And it will be better for the information dissemination related tasks.